\documentclass[twocolumn]{aastex62}
\usepackage{graphicx}	
\usepackage{amsmath}	
\usepackage{amssymb}	
\usepackage{verbatim}
\usepackage{color}

\submitjournal{ApJ}

\shorttitle{ASW pore surface area}
\shortauthors{He et al.}

\begin{document}

\title{The effective surface area of amorphous solid water measured by the infrared absorption of carbon monoxide}

\correspondingauthor{Jiao He}
\email{jhe08@syr.edu}

\correspondingauthor{Gianfranco Vidali}
\email{gvidali@syr.edu}

\author[0000-0003-2382-083X]{Jiao He}
\affiliation{Physics Department, Syracuse University, Syracuse, NY 13244, USA}
\affiliation{Current address: Sackler Laboratory for Astrophysics, Leiden Observatory, Leiden University, PO Box 9513, 2300 RA Leiden, The Netherlands}

\author[0000-0001-8209-2989]{Aspen R. Clements}
\affiliation{Department of Chemistry, University of Virginia, Charlottesville, VA 22903, USA}

\author{SM Emtiaz}
\affiliation{Physics Department, Syracuse University, Syracuse, NY 13244, USA}

\author{Francis Toriello}
\affiliation{Physics Department, Syracuse University, Syracuse, NY 13244, USA}

\author{Robin T. Garrod}
\affiliation{Departments of Chemistry and Astronomy, University of Virginia, Charlottesville, VA 22903, USA}

\author[0000-0002-4588-1417]{Gianfranco Vidali}
\affiliation{Physics Department, Syracuse University, Syracuse, NY 13244, USA}

\begin{abstract}
The need to characterize ices coating dust grains in dense interstellar clouds arises from the importance of ice morphology in facilitating the diffusion and storage of radicals and reaction products in ices, a well-known place for the formation of complex molecules. Yet, there is considerable uncertainty about the structure of ISM ices, their ability to store volatiles and under what conditions. We measured the infrared absorption spectra of CO on the pore surface of porous amorphous solid water (ASW), and quantified the effective  pore surface area of ASW. Additionally, we present results obtained  from a Monte Carlo model of ASW in which the morphology of the ice is directly visualized and quantified. We found that 200 ML of ASW annealed to 20 K has a total pore surface area that is equivalent to 46 ML. This surface area decreases linearly with temperature to about 120 K. We also found that (1) dangling OH bonds only exist on the surface of pores; (2) almost all of the pores in the ASW are connected to the vacuum--ice interface, and are accessible for adsorption of volatiles from the gas phase; there are few closed cavities inside ASW at least up to a thickness of 200 ML; (3) the total pore surface area is proportional to the total 3-coordinated water molecules in the ASW in the temperature range 60--120 K.  We also discuss the implications on the structure of ASW and surface reactions in the ice mantle in dense clouds.

\end{abstract}


\section{Introduction}
In dense clouds in the interstellar medium (ISM), dust grains are covered by
frozen molecules, mostly water ice in the amorphous form \citep{Hagen1981}.
The ice provides a catalytic surface where atoms and molecules are stored and
where reactions leading to the formation of many molecular species take place
\citep{Herbst2009,Vastel2014}. Thus, they are important molecular factories,
together with gas-phase reactions. Laboratory measurement of the surface area
available for catalysis is therefore important for understanding the chemistry
on and in the ice mantle.

In typical laboratory experiments under zero pressure, two recognized forms of
amorphous solid water (ASW) can be formed from water vapor deposition---porous
and non-porous (compact). Whether the structure of ASW is porous or compact
depends on the deposition methods
\citep{Stevenson1999,Kimmel2001a,Dohnalek2003,Raut2007charact}. Generally, lower
deposition temperature and higher deposition angle respect to surface normal favor a higher porosity. If
ASW is grown from water vapor deposition onto a substrate at 130 K or above, the
ice is compact. It was also reported that ASW grown from a collimated beam
of water vapor at normal incidence forms a compact structure even at lower temperatures \citep{Kimmel2001a}.
Omnidirectional deposition of water vapor when the substrate is at lower than 130
K forms porous ASW. Upon heating, porous ice gradually transforms into
non-porous ice. Pore collapse during thermal annealing
\citep{Bossa2012} and as a result of irradiation with ions or UV light
\citep{Palumbo2006} has been previously studied in the laboratory.
However,  a question still remains whether the ice mantle covering dust grains is porous or compact.

One useful signature of porous ASW is the
presence of OH dangling bonds (dOH). It is established that there are two
types, for doubly and triply coordinated water molecules at the ice surface
\citep{Buch1991,Devlin1995}. Their presence is uncovered in the IR tail of the
OH stretch, at 3720 cm$^{-1}$ and 3696 cm$^{-1}$ for doubly and triply
coordinated water molecules, respectively. It is conceivable to link the
presence and strength of the dOH to the porosity of ice; however, this linkage
has been proven difficult to establish unambiguously. Experimental
studies \citep{Palumbo2006, Raut2007compaction, Isokoski2014, Mitterdorfer2014}
show that the total number of dangling bonds is not proportional to the
porosity, and some porosity is retained when the signature of dangling bonds
disappears. This is an important point, since dOH IR signatures have not been
seen in observations so far \citep{Keane2001}. It is also known that the
position and strength of dOH dangling bonds are affected by the presence of
other atoms or molecules (see \citet{He2018b} for a recent investigation of
change in the IR bands of the dangling bonds due to adsorption of H$_2$, D$_2$,
Ar, CO, N$_2$, CH$_4$, and O$_2$). Furthermore, the thermal treatment of ASW
irreversibly changes the network of pores: as the temperature is increased, the
ice morphology changes and pore collapse occurs. In this work, we investigate
again the relation between dOH bands and porosity, and hope to find new insights
into this decades-old problem.

Compared to the studies mentioned above, which mostly focused on measuring the
porosity (or equivalently the density) of the ASW, fewer details are available
about the link between morphology and catalytic properties of ices.
\citet{Raut2007compaction} performed energetic ion bombardment of ASW and found
that the surface area of porous ice decreases at a faster rate than the pore
volume during ion-induced compaction. The underlying reason for this difference
is still not well understood, but several mechanisms have been proposed,
including coalescence of micropores, preferential destruction of smaller pores,
and smoothing of pore wall topology \citep{Raut2007compaction}. Prior
laboratory measurements of porosity based on density \citep{Bossa2014,
Cazaux2015} do not reflect the true catalytic potential of the ASW surface. It
is important to measure the pore surface area that is accessible for the
adsorption of volatiles from the gas phase. \citet{Palumbo2006} studied the
accessible pore surface area after compaction of the ASW by energetic ions.
However, in highly shielded clouds, thermal processes should dominate over
energetic processing, and the temperature dependence of the pore surface area
is the most important. One of the main goals of this study is to fill this gap and
use the infrared absorption spectrum of CO as a tool to quantify the
temperature dependence of the catalytic surface area that is accessible by
volatile molecules condensed from the gas phase.

ASW is also the main component of comets. Although it is widely accepted that
comets are among the most pristine ice objects in the solar system, little is
know about the structure of the ASW in the cometary core. As the structure of
the ice changes when the comet is exposed to solar irradiation, molecules can
be trapped in the ice well beyond the temperature at which they would desorb if
they were adsorbed on the surface \citep{Bar-Nun1985,Smith1997,May2013}.
Recently, the Rossetta mission has detected a number of molecules from comet
67P/Churyumov–Gerasimenko. Notably, molecular oxygen has been detected at an abundance of 4\%
respect to water \citep{Bieler2015}. A satisfactory explanation is still
lacking. To quantify the releasing of volatiles from the comet, it is important
to understand the link between the trapping of volatiles and the ice structure.
In this study we also look into the trapping of CO in ASW under different
temperature conditions.

\section{Experimental setup}
\label{sec:exp}
A detailed description of the apparatus can be found in previous publications
\citep{He2018a, He2018b}, and here only the main features that are relevant to
this study are summarized. Experiments in this study were carried out using
an ultra-high vacuum (UHV) apparatus with a base pressure of $4\times10^{-10}$
Torr. A gold coated copper disk attached to the cold head of a closed-cycle
helium cryostat was used as the sample disc onto which ices were grown. The
temperature of the sample can be controlled between 5 and 350 K to an accuracy
better than 50 mK. Ices are grown by vapor deposition from the chamber
background. CO gas and water vapor entered the chamber via two separate
precision leak valves, which are automated by stepper motors controlled with
LabVIEW programs. Ice thickness in monolayer (ML) are calculated from the
impingement rate based on an integration of the chamber pressure
\citep{He2018b}. One monolayer is defined as $10^{15}$ molecule cm$^{-2}$ on a flat
surface. The ion gauge correction factor and velocity of both water and CO were
taken into account in the calculation. The relative uncertainty in CO dose and
water dose are less than 0.1\% and 1\%, respectively. The main source of
uncertainty is from the hot cathode ion gauge, which has absolute uncertainty
up to 30\%. More details of the deposition control are reported in
\citet{He2018b}. Ice on the sample is measured using a Fourier Transform
InfraRed (FTIR) spectrometer in the Reflection Absorption InfraRed Spectroscopy
(RAIRS) configuration.

In the experiments, we first deposited 200 ML of porous ASW  when the surface was
at 10 K, then heated the ice at a ramp rate of 3 K/minute from 10 K to 200 K.
The RAIR spectra were measured continuously during the heating and we
monitored how the two dOH bands change with temperature. Next, we carried out
a whole set of experiments of CO deposition on top of 200 ML of ASW annealed at
different temperatures. The ASW samples were grown when the sample was at 10 K,
and then annealed at 20, 40, 60, 80, 100, 120, and 140 K for 30 minutes.
Afterwards, the ice sample was cooled down to 20 K (except for the 20
K annealing) before depositing CO continuously until the ASW pore surface was
fully covered by CO, as indicated by the emergence of the longitudinal optical
(LO) mode of CO at $\sim2143$ cm$^{-1}$. The CO deposition rate was chosen so
that there are enough data points during the deposition, and it varies between
experiments. The CO deposition temperature was chosen to be 20 K, because at
this temperature, CO has enough mobility on the surface of p-ASW
\citep{He2018b}.

\section{Modeling}
\label{sec:model}
Ice simulations are conducted using the off-lattice microscopic kinetic Monte Carlo model {\em MIMICK} (Model for Interstellar Monte Carlo Ice Chemical Kinetics), adapted from the works of \citet{Garrod2013} and \citet{Clements2018}, with flat geometry and periodic boundary conditions. The model allows the diffusion of individual molecules to be traced over time, at various temperatures. As well as thermal diffusion (hopping) between surface potential minima, non-thermal diffusion is also allowed, immediately following the deposition of each water molecule onto the surface; the gas-phase translational energy of the molecule and the energy it gains as it enters the surface potential allow it to diffuse if its energy is sufficient to overcome the local diffusion barrier(s). The model uses isotropic Lennard-Jones potentials, which were parameterized within the model by \citet{Clements2018} using experimental density data from the literature for amorphous water formed through background deposition at various temperatures \citep{Brown1996}.

In the present models, water is deposited at interstellar temperatures (10 to 20 K) and then heated at laboratory rates (1 to 3 K min$^{-1}$) up to a temperature of 150 K. First, the water molecules are deposited using background deposition onto a square surface of length 650 {\AA}, significantly smaller than a surface used in the experiments for computation time.
A deposition rate of 10$^{13}$ cm$^{-2}$ s$^{-1}$ and temperature of 10 K were used and two thicknesses were tested (25 ML and 200 ML). Surface area and density are calculated for each ice and measured during heating.

The ice surface area (including pore surfaces) is calculated by counting the number of surface molecules. This value is then divided by the total number of water molecules in the ice. This ratio corresponds to the coverage of the surface to the total ice thickness in monolayers. With the microscopic model the surface coverage can be directly measured. An average of the thin ice (25 ML) and the thick ice (200 ML) was averaged to calculate the pore surface, as we later discuss the surface area is dependent of thickness between 10 to 200 ML.
Images were created using the freeware POV\textendash Ray to visualize the entire ice or, using cross sections,  the connectedness of the pores.

\section{Results and Analysis}
\label{sec:results}
\subsection{Infrared characterization of pure ASW}
\begin{figure}[ht!]
\plotone{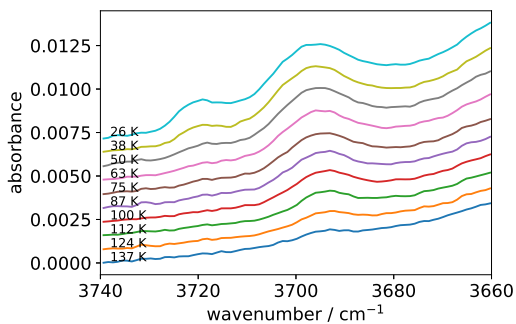}
\caption{RAIR spectra of 200 ML water ice during heating at various
temperatures. The water ice is deposited from the background when the surface
is at 10 K. The heating ramp rate is 3 K/minute. Spectra are offset for
clarity.\label{fig:doh_stack}}
\end{figure}

\begin{figure}[ht!]
\plotone{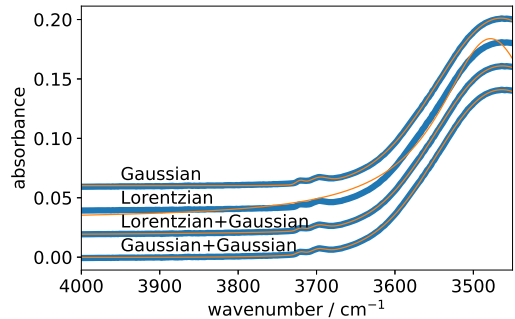}
\caption{Example of fitting of the RAIR spectrum of bulk water OH stretching
mode absorption using different fitting schemes. The small features of the
dangling OH (dOH) bonds located at $\sim$3696 cm$^{-1}$ and $\sim$3720
cm$^{-1}$ are each fitted with a Gaussian function. The left side of the main peak
is fit using the four schemes labeled in the figure. \label{fig:fit_compare}}
\end{figure}

\begin{figure}[ht!]
\plotone{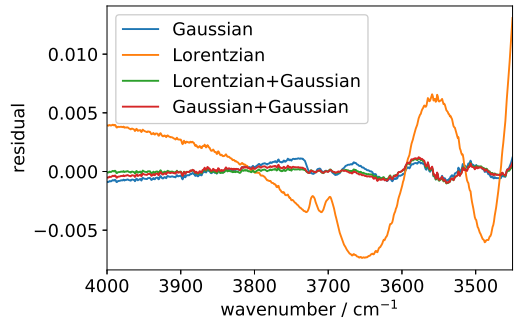}
\caption{Residual of the fittings in Figure~\ref{fig:fit_compare}.
\label{fig:residual}}
\end{figure}

\begin{figure}[ht!]
\plotone{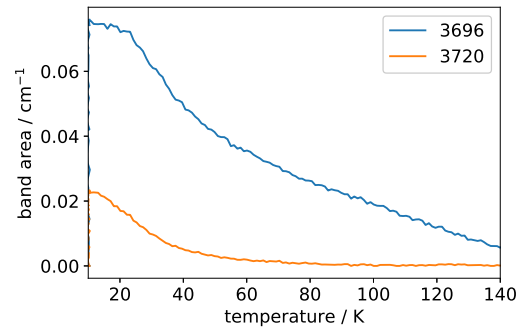}
\caption{The area of the two dOH absorption bands during warming up of a 200 ML
water ice grown at 10 K and heated at 3 K/minute. \label{fig:wat_doh_area}}
\end{figure}

ASW has three main absorptions in the mid-infrared region: OH stretching at
$\sim3300$ cm$^{-1}$, bending mode at 1640 cm$^{-1}$, and libration mode at
$\sim700$ cm$^{-1}$. On the blue shoulder of the OH stretch band, there are two
small absorption features at $\sim3696$ cm$^{-1}$ and $\sim3720$ cm$^{-1}$,
generally attributed to 3-coordinated and 2-coordinated water molecules, respectively
\citep{Buch1991}. The dOH bands contain important information regarding the
structure of the ASW, and are the focus of this subsection. The dOH region of
the RAIR spectra of the 200 ML ASW during heating is shown in
Figure~\ref{fig:doh_stack}. As the ice temperature is raised, both dOH bands
decrease. By 60 K, the $3720$ cm$^{-1}$ band is almost gone, while the $3696$
cm$^{-1}$ band persists until above 140 K. To quantify the temperature
dependence of both bands, we use two Gaussian functions to fit the two dOH
bands. Because the two dOH bands lie on the tail of the main OH stretching band,
it is important to find  a function that fits the baseline. In previous
studies, spline interpolation or polynomial functions were typically employed to fit
the baseline \citep{Dartois2013, Bu2016, Mitchell2017}. We find that for the
work presented here they are not good enough for an accurate description of the
baseline, and lead to inaccuracies in dOH band area calculations. Gaussian and
Lorentzian functions are very often used to fit solid state infrared absorption
features. Typically, disordered ices have relatively broad Gaussian lineshapes,
while crystalline ices have narrower Lorentzian lineshapes. We tried (1) one
Gaussian; (2) one Lorentzian; (3) two Gaussians; and (4) one Gaussian and one
Lorentzian functions to fit the blue side of the OH stretch peak; the two dOH
bands are also included in the fitting. Figures ~\ref{fig:fit_compare} and
~\ref{fig:residual} show the fitting and the residuals, respectively. It can be
seen that the fitting using one Gaussian and one Lorentzian, in addition to two
Gaussians for the dOH bands achieve the best results. For analyses that do not
require high accuracy, one Gaussian function also fits the blue half of the OH
stretch well. In the remaining of this work, we use one Gaussian and one
Lorentzian to fit the OH stretch band. Based on the above fitting scheme, the
band areas of both dOH bands during warming of the 200 ML ASW are calculated
and presented in Figure~\ref{fig:wat_doh_area}. At $\sim$60 K, the $3720$
cm$^{-1}$ band becomes negligible, which suggests the disappearance of 2-coordinated dangling bonds. This is consistent with previous experimental
studies \citep{Raut2007compaction, Smith2009, Bu2015}. The $3696$ cm$^{-1}$ band drops
linearly with temperature from 60 K to above 140 K. The residual $3696$ cm$^{-1}$ band at above 140 K is largely due to the dangling bonds located on the outer surface \citep{Smith2009}.

\subsection{CO on ASW}
\label{sec:co_on_asw}
In the next set of experiments, we use the infrared bands of CO to probe the
pore surface area of ASW annealed at different temperatures. It is well-established that the infrared absorption feature of CO interacting with water ice is different from that of pure CO ice. The RAIRS of pure CO shows the longitudinal
optical (LO) mode at 2143 cm$^{-1}$ (which is the typical one that is excited
in the grazing mode geometry) while CO interacting with water shows two bands
at $\sim$2140 cm$^{-1}$ and $\sim$2152 cm$^{-1}$. According to our previous
laboratory measurement of the diffusion of CO on the surface of porous amorphous solid water (p-ASW, \citet{He2018b}), diffusion of CO becomes significant at about 15 K. At 20 K,
the diffusion is very efficient. If CO is deposited on top of ASW at 20 K, CO
should diffuse into the pores and occupy the pore surface of the ASW. Once the
whole surface area is covered by CO, and CO begins to build up as ``pure'' CO
ice, the LO mode emerges. By examining the amount of CO deposited at which the
LO peak emerges, the accessible surface area of p-ASW can be obtained. In the
following, we first present a detailed analysis of the results of CO deposition
on ASW annealed at 60 K and cooled to 20 K, and then show the results at other
annealing temperatures.

\begin{figure}[ht!]
\plotone{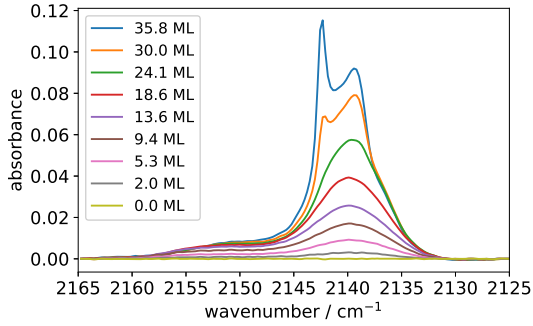}
\caption{The RAIR spectra of CO deposited on top of 200 ML ASW that is annealed
at 60 K for 30 minutes and cooled down to 20 K. The CO dose for each spectrum
is shown in the inset. \label{fig:spec_60K}}
\end{figure}

Figure~\ref{fig:spec_60K} shows the RAIR spectra of C-O stretching mode during
deposition of CO on ASW annealed at 60 K and cooled to 20 K. At low CO
deposition doses, there are two broad components centered at $\sim$2140
cm$^{-1}$ and 2152 cm$^{-1}$. When the CO dose is over $\sim$30 ML, the LO mode
at 2143 cm$^{-1}$ emerges, which we take as a sign of full coverage of the pore
surface.

There have been several experimental studies of the interaction between CO and
ASW surface \citep[e.g.][]{Fraser2004,Collings2005}. Although It is generally
accepted that the $\sim$2152 cm$^{-1}$ component is due to the adsorption of CO
on the dOH sites of ASW, there is no direct experimental evidence, as far as we
know, that demonstrates the correlation between dOH bands and the 2152 cm$^{-1}$
component. We used two Gaussian functions to fit the 2140 cm$^{-1}$ and 2152
cm$^{-1}$ components, and one Lorentzian function to fit the 2143 cm$^{-1}$
component, and then we studied how these three components change with CO
deposition dose. An example of fitting is shown in Figure~\ref{fig:60K_fit}. We
apply similar fitting to all of the spectra in this experiment. The resulting
band areas for the three components are shown in Figure~\ref{fig:60K_peakareas}.

\begin{figure}[ht!]
\plotone{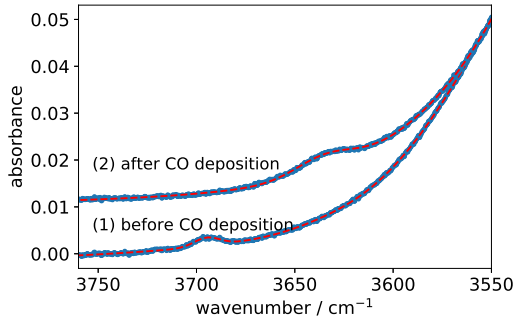}
\caption{The dOH region of the RAIR spectra of 200 ML water ice annealed at 60
K and cooled down to 20 K (1) ; and after 21 ML of CO deposition (2). Dashed lines are
the fitting. Spectra are offset for clarity. \label{fig:60K_dOH}}
\end{figure}

\begin{figure}[ht!]
\plotone{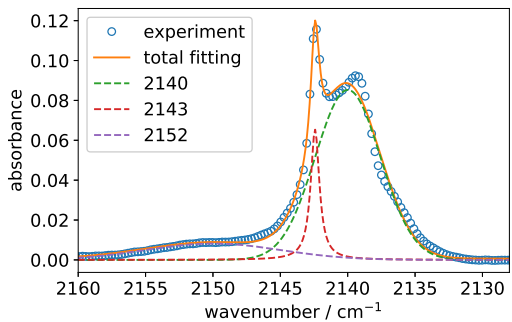}
\caption{An example fitting of the spectra in Figure~\ref{fig:spec_60K} using
two Gaussian functions and one Lorentzian function. \label{fig:60K_fit}}
\end{figure}

\begin{figure}[ht!]
\plotone{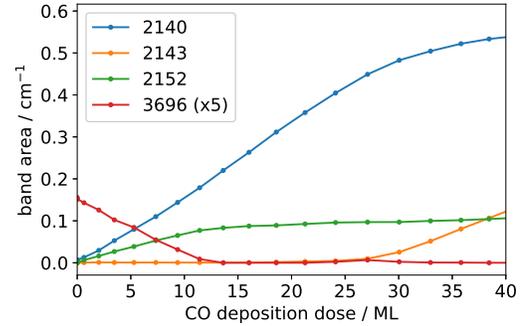}
\caption{The band area of the three components of the CO absorption profile and
the dOH band at 3696 cm$^{-1}$ during CO deposition on 200 ML ASW annealed to 60 K.
Fittings are done as shown in Figure~\ref{fig:60K_fit}
\label{fig:60K_peakareas}}
\end{figure}

As was discussed in \citet{He2018b}, introducing CO in ASW shifts the dOH
bands. In Figure~\ref{fig:60K_dOH}, the dOH region of the RAIR spectra before
and after the CO deposition is shown. Before CO deposition, the dOH band is at
3694--3696 cm$^{-1}$ (the peak position varies between 3694 and 3696 cm$^{-1}$, depending on the annealing temperature; hereafter we refer to this peak as the  the 3696 cm$^{-1}$ peak), while after CO deposition, the area of
the 3696 cm$^{-1}$ peak decreases to zero and the dOH induced by CO shows up at
$\sim$3636 cm$^{-1}$. We used one Gaussian function to fit the 3696 cm$^{-1}$
peak and one Gaussian function to fit the 3636 cm$^{-1}$ peak, and obtained how
the two peaks change with increasing CO deposition. The area of the 3696 cm$^{-1}$ peak
is shown in Figure~\ref{fig:60K_peakareas}, together with the peak areas of the
three components of C-O stretching mode. Between 0 and 12 ML, the
3696 cm$^{-1}$ band area decreases to zero. At the same time, the band area of
2152 cm$^{-1}$ component increases from 0 to the saturation level. The
anti-correlation between these two bands is evident. This is direct evidence
that the 2152 cm$^{-1}$ component is associated with CO binding to the dOH
bonds.

At about 27 ML of CO deposition, the 2140 cm$^{-1}$ band begins to saturate,
while at the same time the 2143 cm$^{-1}$ LO band emerges. This demonstrates
that at about 27 ML of CO deposition, all the pore surface area is occupied,
and ``pure'' CO starts to build up. This happens at a higher CO dose than the full covering of the dOH bonds, likely because CO molecules preferentially occupy the dOH sites than non-dOH sites. In a prior study by \citet{Zubkov2007}, it is reported that the full coverage of pore surface by nitrogen adsorption happens simultaneously with the saturation of the shifted dangling bond intensity. They suggested that N$_2$ does not preferentially bind to dangling OH groups. The difference between that
work and this one is possibly due to the relative interaction energies of the two adsorbates. While nitrogen adsorption shifts the 3-coordinated dangling OH peak to 3668 cm$^{-1}$, CO adsorption shifts it by a larger amount, to 3635 cm$^{-1}$ \citep{He2018b}.
We take the CO dose at which 2143
cm$^{-1}$ component just starts to show up, 27 ML in this case, to be the pore
surface area. The fact that the 3696 cm$^{-1}$ band disappears after CO
deposition indicates that all of the pore surface area is accessible for CO
adsorption, and there is insignificant number of closed cavities inside the ASW (which would have been detected by residual dOH bonds). Therefore 27 ML is
the accessible area and is also the total pore surface area of the 200 ML ASW
annealed to 60 K.

\begin{figure*}
\gridline{\fig{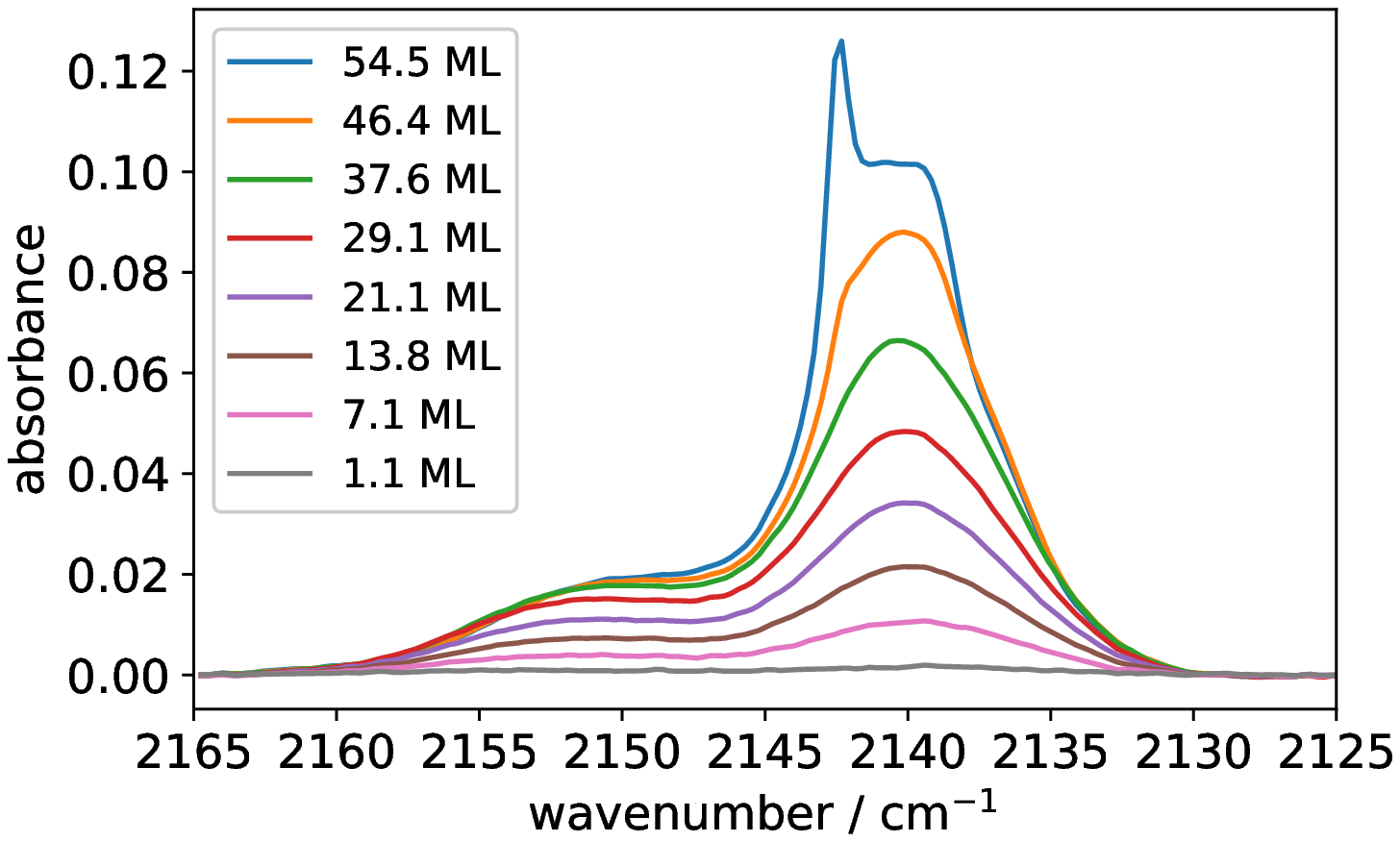}{0.33\textwidth}{20 K}
          \fig{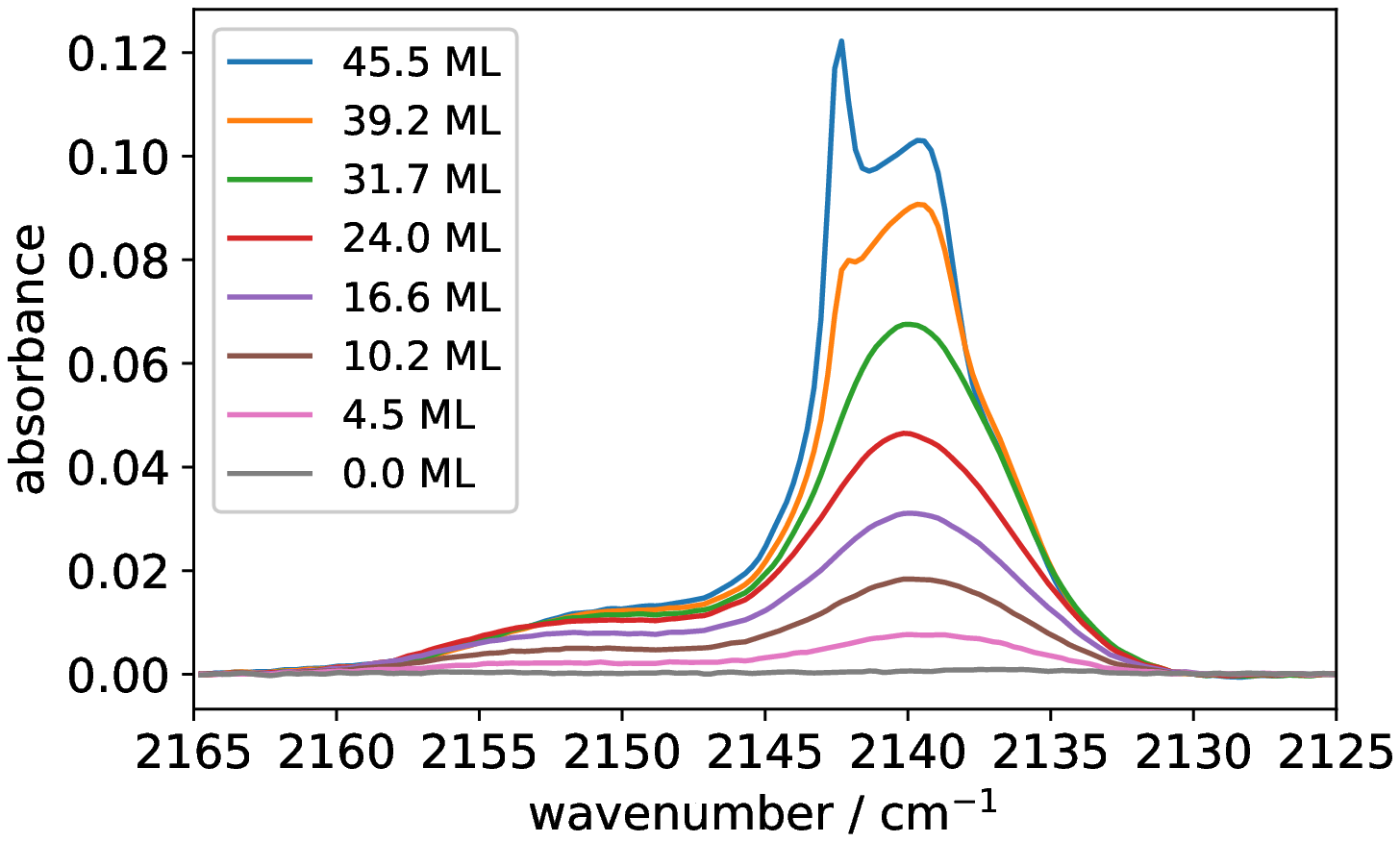}{0.33\textwidth}{40 K}
          \fig{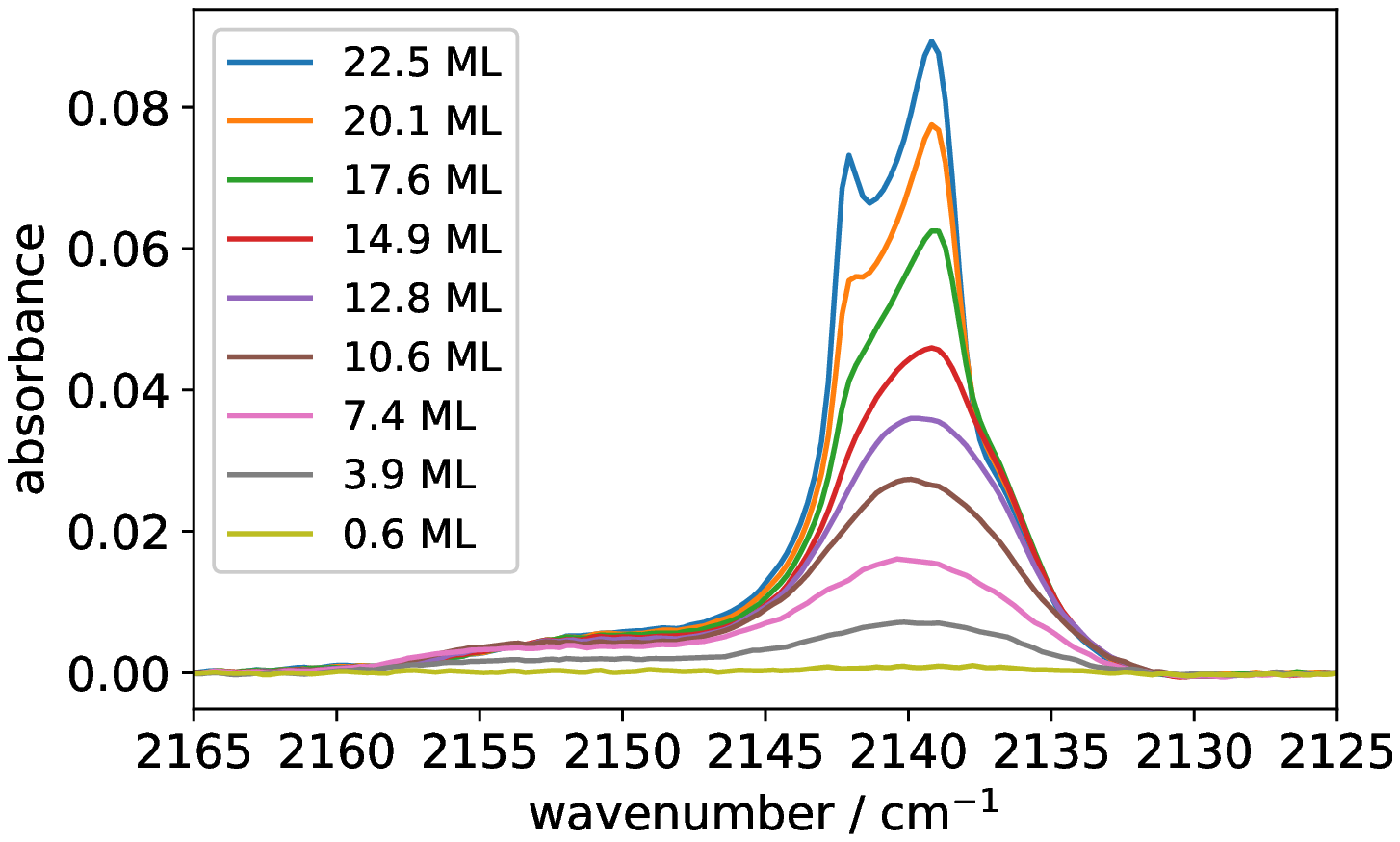}{0.33\textwidth}{80 K}
          }
\gridline{\fig{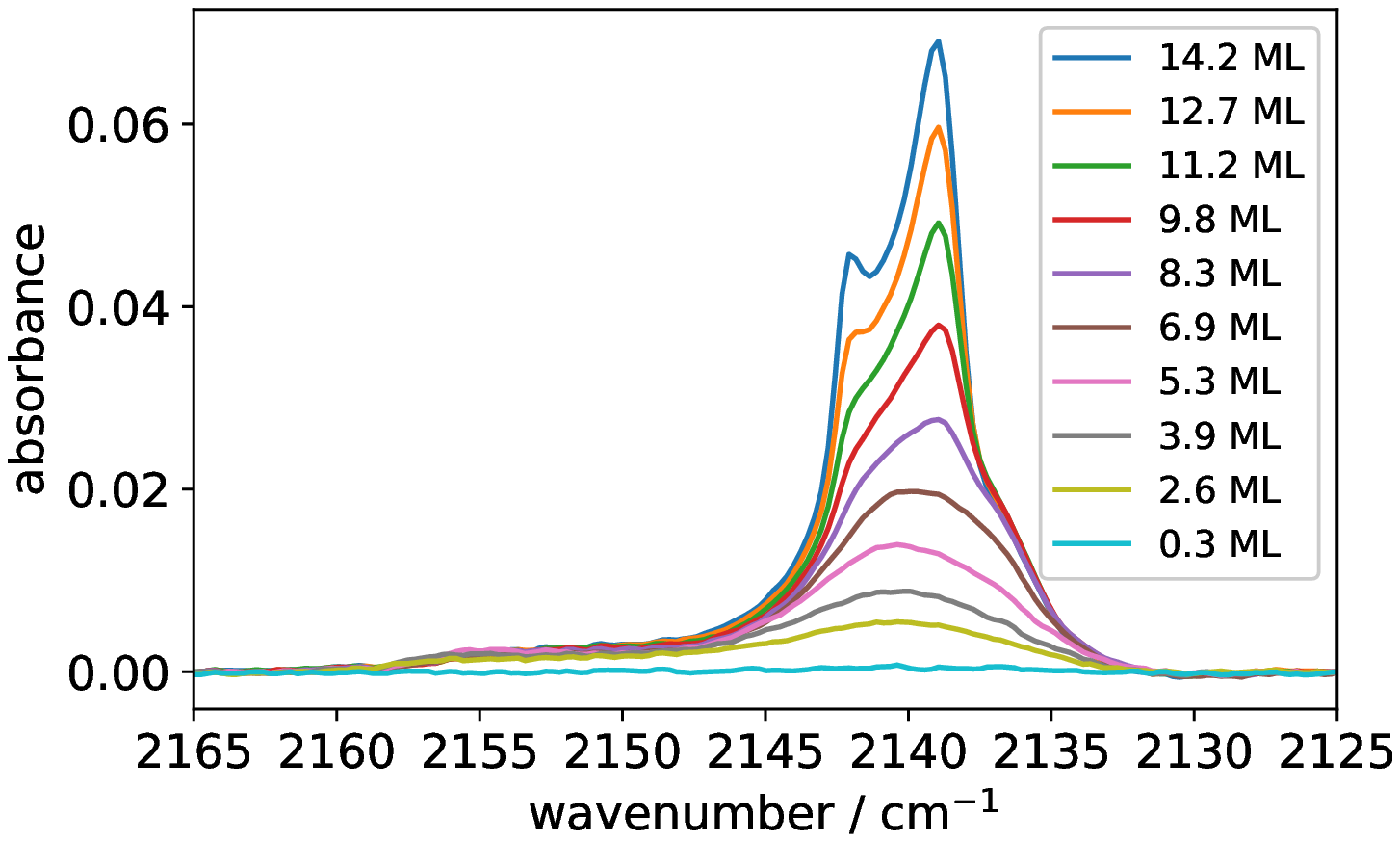}{0.33\textwidth}{100 K}
          \fig{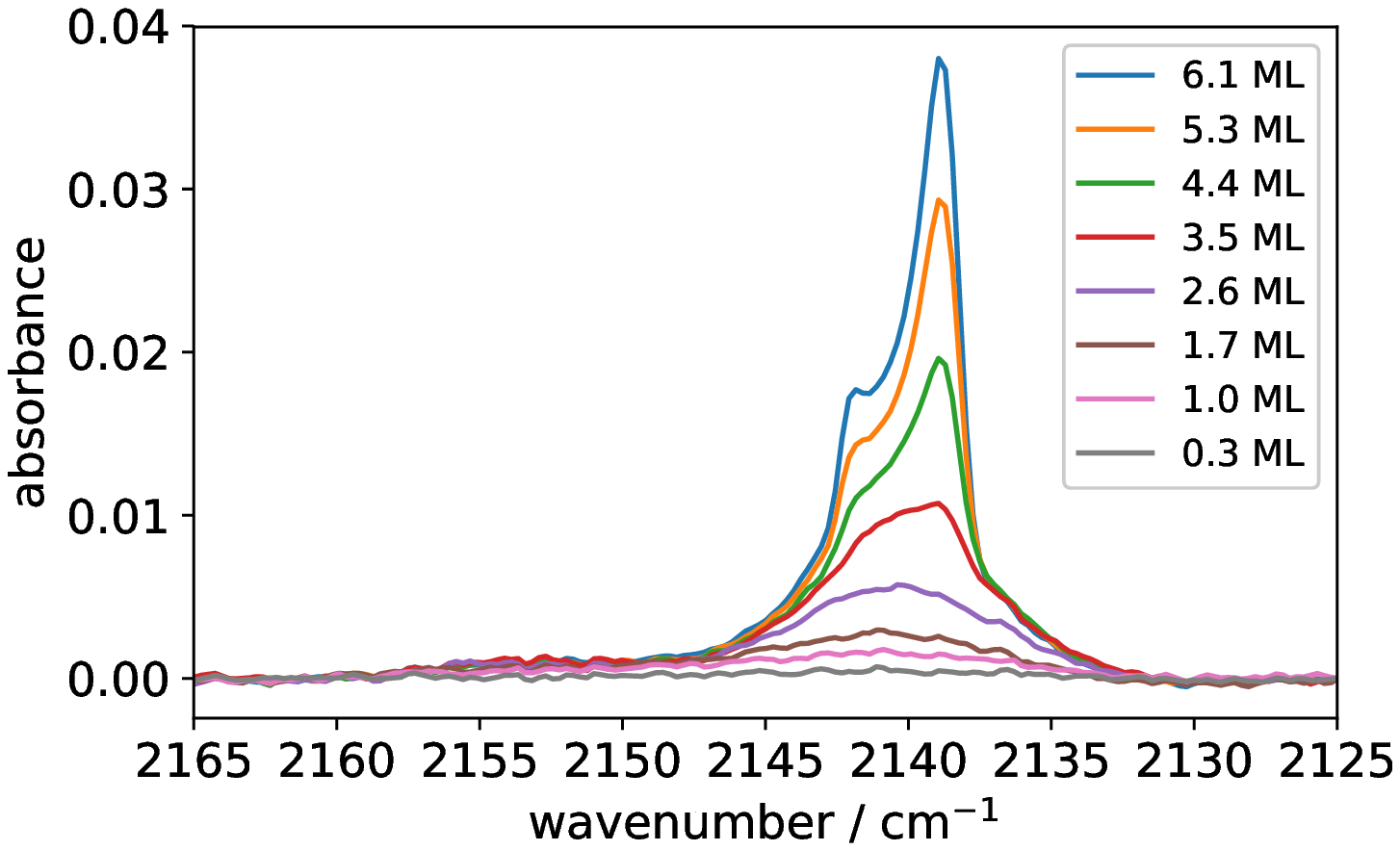}{0.33\textwidth}{120 K}
          \fig{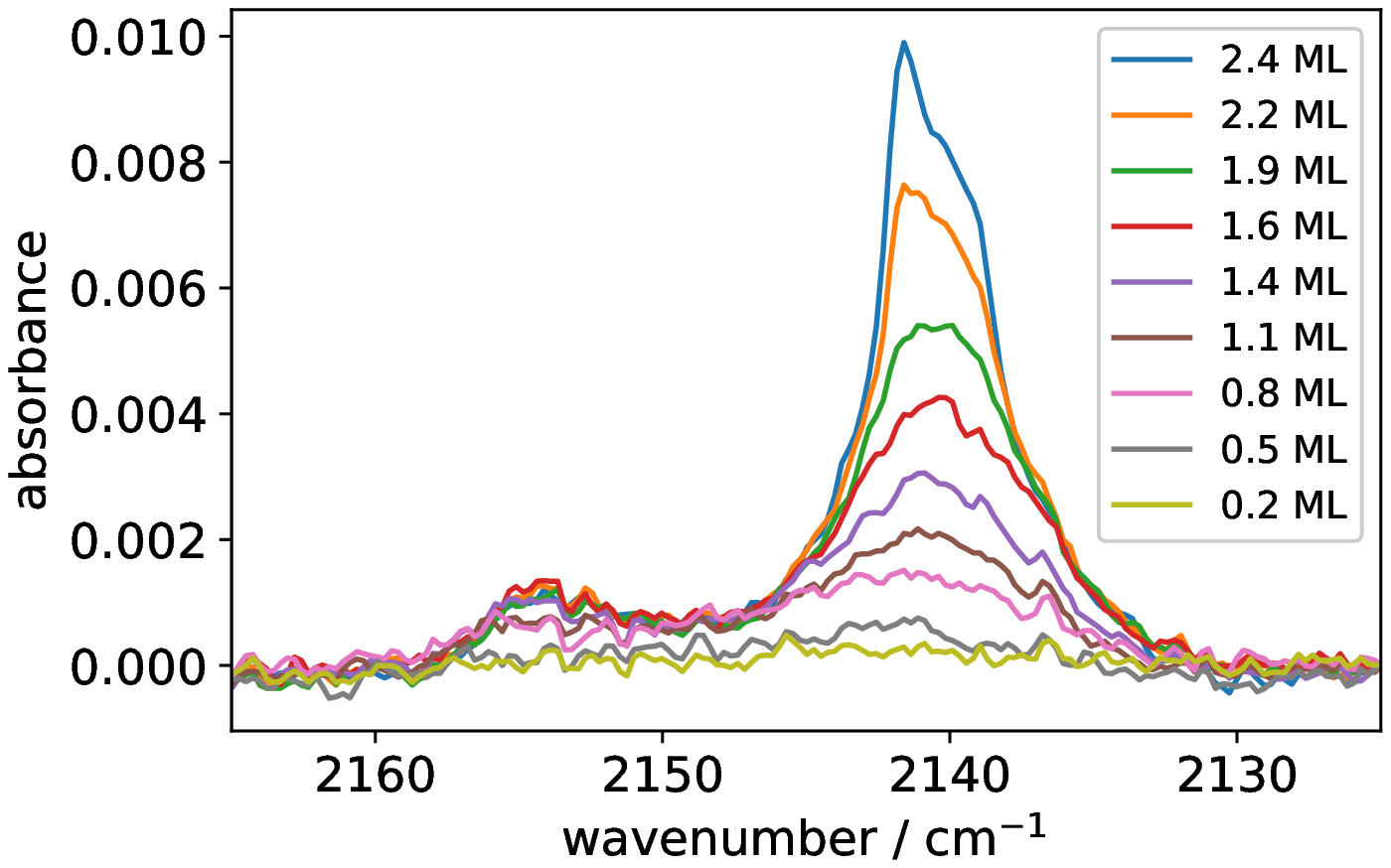}{0.33\textwidth}{140 K}
          }
\caption{The RAIR spectra of CO deposited on top of 200 ML ASW that is annealed
at 20, 40, 80, 100, 120, and 140 K, and cooled down to 20 K. The CO dose for
each spectrum is shown in the inset.\label{fig:specs}}
\end{figure*}

\begin{figure}[ht!]
\plotone{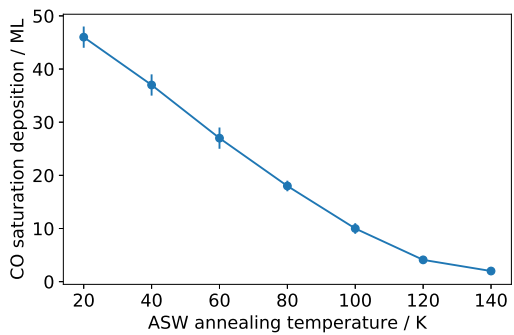}
\caption{Accesible pore surface area in 200 ML of ASW that are annealed at
different annealing temperatures. The pore surface area is measured by the
amount of CO that fully covers the pore surface. \label{fig:sat_area}}
\end{figure}

Similar CO depositions were carried out on 200 ML ASW samples that were
annealed at 20, 40, 80, 100, 120, and 140 K, and cooled down to 20 K. RAIR
spectra were recorded during CO depositions at 20 K and are shown in
Figure~\ref{fig:specs}. We determine the pore surface areas for the ASW
annealed at different temperatures by visually examining the CO deposition dose
at which the 2143 cm$^{-1}$ component emerges. The ASW surface area versus
annealing temperature is shown in Figure~\ref{fig:sat_area}. The pore surface
area decreases linearly with annealing temperature almost up to 120 K, above
which the surface area becomes about 2 ML. Considering that the surface of ASW is
rough, 2 ML covers probably the very top of the surface, i.e., the
ice--vacuum interface. The ice becomes fully compact at 140 K. The almost linear decrease in pore surface area with annealing temperature in the range 20--120 K is also seen in the neutron scattering experiments by another group (Sabrina G\"{a}rtner, private communication).

\subsection{Trapping of CO in ASW}
\begin{figure}[ht!]
\plotone{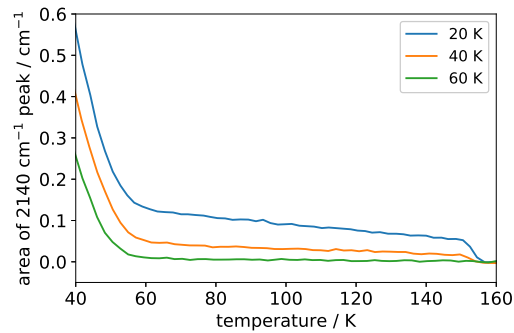}
\caption{Band area of the 2140 cm$^{-1}$ peak during warming up of (1) CO
adsorbed on ASW that is annealed at 20 K; (2) CO adsorbed on ASW that is
annealed at 40 K and cooled down to 20 K; (3) CO adsorbed on ASW that is
annealed at 60 K and cooled down to 20 K. \label{fig:trapping}}
\end{figure}

In the previous section, we focused on the infrared spectra during CO depositions. Here in this section, we focus on the TPD stage of the same set of experiments. After the deposition of CO on ASW at 20 K, the ice was heated up from 20 K to 200
K at a ramp rate of 0.1 K/s while RAIR spectra were measured continuously. The
band area of the C-O stretching mode was calculated for each spectrum during
warming up (see Figure~\ref{fig:trapping}). For ices that are annealed at 60~K
and above, the C-O stretch band area becomes zero after the temperature goes past 60 K ( for clarity purpose, curves for 80 K and above are not shown in the figure). This is in agreement with the study by \citet{Horimoto2002} who carried out similar experiments using methane instead of CO. In the figure, the
desorption of CO from the ice can be separated into three regions. The first
region is below about 55 K, which is the temperature at which CO on ASW surface
(including the surface of pores) desorbs. The second region is from about 55
K to about 150 K, during which the CO band area drops linearly with
temperature. These are the CO molecules that are trapped in the ASW matrix and
released back into the gas phase gradually. Here we don't exclude the
possibility that the band strength of CO buried inside bulk ASW can change with
temperature. Indeed, experimental measurements by \citet{Schmitt1989} have
found that the band strength of C-O stretching for CO buried in water ice has
a reversible component that decreases almost linearly with the temperature between
50 K and 120 K. The irreversible component corresponds to the gradual
releasing of CO from the bulk ASW. The third desorption happens when the ASW
crystallizes, and all of the remaining CO molecules are forced out of the ice.
This is sometimes referred to as the ``molecular volcano'' \citep{Smith1997}.
The amount of CO that is in the ice at about 60 K represents the CO that is
trapped inside the ASW matrix, and we define it as the trapping amount. When
the ASW is annealed to 60 K or above, the ASW does not trap any CO. The lower
the annealing temperature, the higher the number of CO molecules that can be trapped. The
linear decrease of C-O stretching band area during heating is similar to that
of CO$_2$ (see Figure 4 of \citet{He2018c}). This suggests that the linearity
may be a general phenomenon that occurs to all volatiles that are trapped in
ASW. In a forthcoming paper, we'll present a detailed study on the trapping of
volatiles in ASW.

\subsection{Dangling OH bonds during CO deposition}
\begin{figure}[ht!]
\plotone{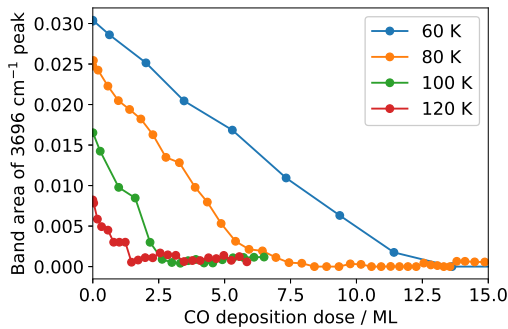}
\caption{Band area of the dOH bond absorption at 3696 cm$^{-1}$ after
deposition of CO at 20 K on 200 ML ASW that has been annealed at 60, 80, 100, and 120 K.
\label{fig:3696area}}
\end{figure}

During CO deposition on ASW, as the pore surface is gradually covered by CO,
the dOH band at 3696 cm$^{-1}$ decreases, and the band at 3636 cm$^{-1}$
increases. We applied the fitting scheme as discussed above to obtain the area
of the dOH band during CO deposition. Figure~\ref{fig:3696area} shows the area
of the 3696 cm$^{-1}$ dOH band during CO deposited on ASW that has been annealed
at different temperatures. For the ASW that was annealed at
140 K, the dOH band area is too small, and is not presented in the figure. For
annealing temperature of 20 K and 40 K, there are two dOH bands after
annealing, and the fitting of the peaks is more complicated and are not
considered here. The main finding from Figure~\ref{fig:3696area} is that the
dOH always drop to zero at high enough CO doses, regardless of the annealing
temperature. This suggests that almost all the pore surface inside the ASW are
accessible to CO, and the pores throughout the whole ice are interconnected.
This agrees with the previous results by \citet{Raut2007charact} which
demonstrated that all of the pores are interconnected and are accessible to
CH$_4$ adsorption.

\subsection{Modeling of ASW Ice Porosity}
\begin{figure*}[ht!]
  \includegraphics[width=\linewidth]{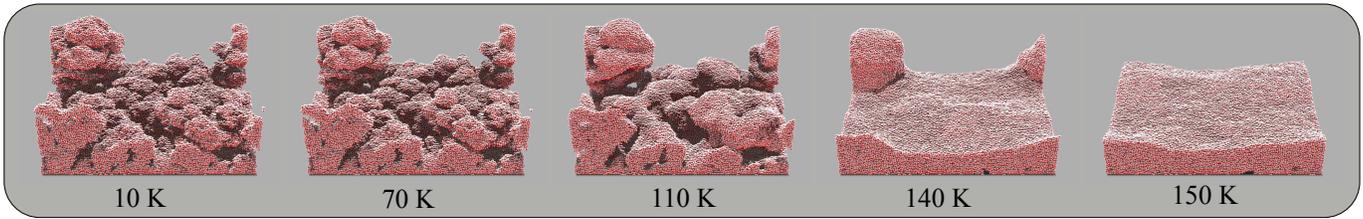}
  \caption{A 25 ML amorphous water deposited and heated at 3 K min$^{-1}$ to 160 K.}
  \label{fig:heatingModel}
\end{figure*}

Figure \ref{fig:heatingModel} shows the structure of simulated water ice; the column-like structure becomes smoother with increasing temperatures, until eventually an entirely smooth structure is obtained at 140 to 150 K. In the model, the initial ice was deposited at 10 K and then heated to $\sim$150 K where the ice  starts to desorb into the gas phase. The first two images of the model (at 10 and 70 K) have essentially the same structure. The model indicates there is little to no re-arrangement of the ice until  the temperature of 60 K is reached. At 60 to 80 K, diffusion of water becomes efficient enough to play a role in the surface area and porosity, and increases at higher temperatures. The structure begins to smooth, by eliminating first the smaller pores until gradually all the pores are removed. Through this process, the ice reaches its maximum density at 150 K. It is important to note the model does not include the phase change from amorphous to crystalline ice, which would occur at $\sim$140 K; this does not alter the results of the model as the main focus is between temperatures of 10 to 140 K.

\begin{figure}[ht!]
\plotone{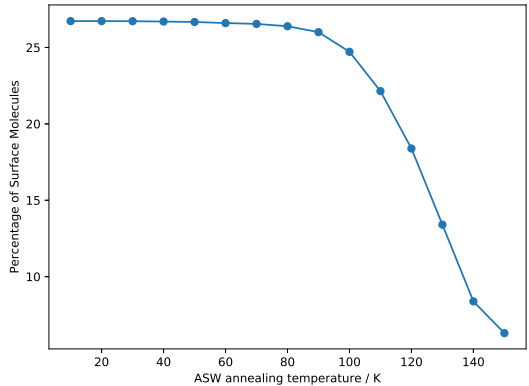}
\caption{Accessible pore surface area in the  model of ASW as it is heated at 1 K min$^{-1}$. The pore surface area is measured by the percentage of surface to total water molecules. The error bars  are calculated by using both 25 and 200 ML model coverages, and are essentially insignificant.
 \label{fig:ModelCov}}
\end{figure}

In the laboratory results shown in Figure~\ref{fig:sat_area}, we see that the accessible surface area decreases steadily up to 140 K. As stated above, the 3-coordinate dOH absorption band (3696 cm$^{-1}$) linearly decreases from 60 to 140 K, likely corresponding to the decrease in the surface area; this matches the decrease of the exposed surface in the model. As seen in Figure \ref{fig:ModelCov}, the coverage steadily decreases after 80 K is reached. Until that temperature is reached, very little rearrangement and pore collapse occur; this is probably due to the fact that the model uses isotropic potentials, and is not sensitive to defects (OH dangling bonds) which the experiment is sensitive to. However, within the model we see a reasonable match at lower temperatures given that CO is a proxy for the extent of the accessible exposed H\textsubscript{2}O network.

\begin{figure}[ht!]
\plotone{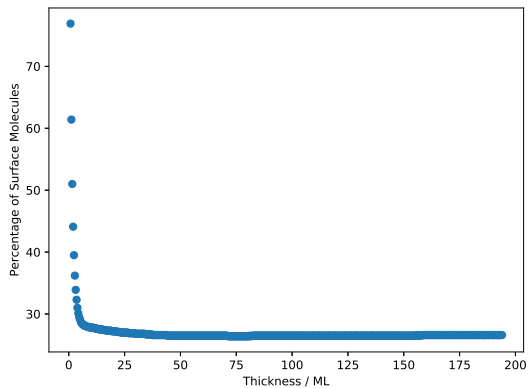}
\caption{Ratio of the number of surface water molecules to the total numbe of water molecules obtained in modeling during deposition of 200 ML water at 10 K.
 \label{fig:areaVSthickness}}
\end{figure}

Figure~\ref{fig:areaVSthickness} shows the modeling results of the ratio of the number of surface molecules to the total number of  molecules during the deposition of 200 ML water onto a 10 K surface. In the first few monolayers, there is a large fraction of surface molecules. After the thickness reaches more than $\sim$10 ML, the fraction of surface molecules is no longer dependent on the thickness. This suggests that the structure of the ASW film is homogeneous and the conclusions in this work based on measurements of 200 ML ASW can be generalized to other thicknesses as well, as long as the ice is thicker than a threshold, in this study, $\sim$10~ML. This is more or less in agreement with previous studies by \citet{Stevenson1999, Kimmel2001a, Kimmel2001b, Smith2009}, although the threshold thickness in those studies differ from this study.

Previously, Kimmel et al. 2001b.\citet{Kimmel2001b} used a kinetic model where a hit and stick method was used. Each individual molecule sticks to the surface being placed depending on the trajectory angle provided. This model does not include kinetic energy, but a parameter that designates how many times each incoming particle is allowed to hop before being permanently sticking. The images presented from Kimmel et al. show that the pores are also interconnected at the temperature of deposition, but does not include a linear warm-up of the ice. The model presented here is a kinetic model and hopping is set by the temperature, which in this case is essential as we linearly increase the temperature to replicate the experimental results. We show that the pores are interconnected and maintain this structure for high temperatures.

Unlike the experiments, within the model no rearrangement occurs below 60 K, because of the isotropic treatment used. Essentially, the model does not show the small scale rearrangement due to re-alignment of water molecules within their original potential. While the model cannot achieve these changes, it can provide a direct way to measure the surface coverage of water ice. The surface area can be monitored during heating instead of requiring CO adsorption experiments where the water ice must be cooled down to 20~K to measure the amount of CO adsorbed on the ASW. Furthermore, the determination of surface area using CO adsorption as in the laboratory experiments may not be exactly the same as that from counting the number of water molecules on pore surfaces as in the modeling. A small difference between these two methods is possible.

\begin{figure}[ht!]
  \includegraphics[width=\linewidth]{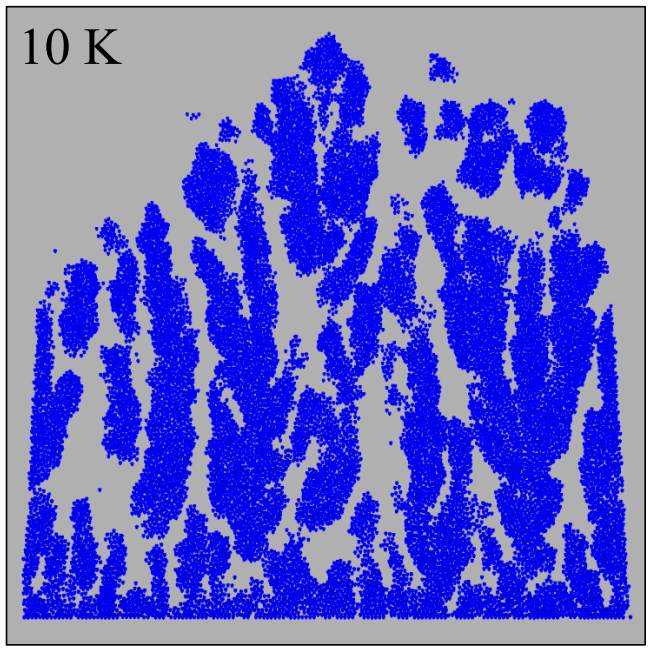}
  \caption{A slice of a 200 ML thick amorphous water deposited at 10 K imaged with POV\textendash Ray. The thicker ice was used to demonstrate the interconnectedness as it was more obvious than in the much thinner ice of 25 ML.}
  \label{fig:sliceMod}
\end{figure}

Figure \ref{fig:sliceMod} shows the interconnectedness of the pores. A portion of ice was imaged to show the inner structure and not the total structure. Visually it shows that most pores are connected within the shown plane. As the ice is heated the pores collapse until eventually empty cavities within the water ice are left.The cavities appear to be the remnants of the initial column-like structure, which minimize their potentials by forming approximately spherical structures. The encapsulated pores are fairly small in size with widths around 2 to 3 nanometers. These cavities may allow entrapment of some volatile species such as CO until a later temperature. By 150 K all cavities have collapsed and the volatiles have either been desorbed or are stuck within a water matrix.

\section{Discussion and Astrophysical Implications}
\label{sec:astro}
One of the main spectroscopic pieces of evidence of porous ASW is the presence
of dOH bonds, which have been seen in numerous laboratory experiments. Whether
the infrared signature of dOH bonds at 3696 cm$^{-1}$ and 3720 cm$^{-1}$ is
a good measurement of porosity has been debated. \citet{Palumbo2006} performed
energetic ion bombardment on ASW and found that the decrease of pore surface
area is four times less than the decrease in dOH absorption. From this
experiment, one might conclude that the dOH band area is not proportional to the
pore surface area. However, ion bombardment also induces chemistry that produces
molecules such as O$_2$, O$_3$, H$_2$O$_2$, which may interact with the dOH
bonds and shift or shield the dOH bands \citep{He2018b}. It is unclear from
their experiments what is the relation between dOH band area and pore surface
area for pure ASW. In this regard, thermal processing of ASW is a more
appropriate experiment. There have been experiments that focused on thermal
processing of ASW \citep{Bossa2012, Bossa2014, Isokoski2014}. However, those
studies quantified the density of the ice instead of the pore surface area. In
this study, we measured the pore surface area of ASW that is annealed at
different temperatures, and also quantified the temperature dependence of dOH
band area during heating of ASW. Additionally, we used a kinetic Monte Carlo model to determine the pore surface area of ASW during heating, by computing directly the total number of surface molecules. We have shown in Figure~\ref{fig:wat_doh_area}
that the 2-coordinated dOH (3720 cm$^{-1}$) decreases sharply between 10
K and 60 K, and disappears almost completely by $\sim$60~K. There is possibly
a smoothing of the pore surface or merger of smaller pores to form larger ones.
The 3-coordinated dOH absorption band (3696 cm$^{-1}$) decreases almost
linearly between 60 and 140 K. This linear decrease with temperature can be
compared with the linear decrease of pore surface area shown in
Figure~\ref{fig:sat_area}. Both seem to decrease linearly with temperature,
although the curve in Figure~\ref{fig:sat_area} drops to the minimum at a slightly
lower temperature than in Figure~\ref{fig:wat_doh_area}.  This small difference
can be explained by the fact that in measuring the pore surface area using CO
adsorption, ASW was annealed for 30 minutes, while when measuring the dOH bands
shown in Figure~\ref{fig:wat_doh_area}, the temperature was ramped up
continuously without the annealing step. The comparison between these two
figures suggests that at least under our experimental conditions, the infrared
absorption band of 3696 cm$^{-1}$ seems to be a good measurement of the
pore surface area in the temperature range from 40 K to 140 K.

The presence of cavities inside ASW has been reported or mentioned in several
prior studies \citep{Eldrup1985, Horimoto2002, Zheng2007}.  However, it is
unclear whether these cavities are closed inside the bulk ASW or interconnected
and accessible from the vacuum--ice interface. In Figure 3 of
\citet{Raut2007compaction} and Figure 8 of \citet{Cazaux2015}, it was hinted
that there are closed cavities, but there was no discussion about the
connectivity of the cavities. In Figure~\ref{fig:3696area} in this work, it is
evident that after CO adsorption, the 3696 cm$^{-1}$ band always drops to zero,
regardless of the annealing temperature. This suggests that there is an insignificant number of closed cavities inside the bulk ASW, and almost all of the cavities (pores) are connected to the vacuum--ice interface. This is verified by the modeling, which shows that volatile species should indeed be able to access almost all of the pore surface inside the ice. This is also in agreement with the experimental results of \citet{Raut2007charact}, who found that the number of closed pores is insignificant in a 1000 ML ASW. However, we have to point
out that this conclusion may not be applicable to an ASW much thicker than 1000
ML. \citet{Bu2016} reported that a thick ASW may crack spontaneously during growth
or during warming up. It remains a question how the spontaneously cracking
affects the connectivity of the cavities/pores inside ASW.

From Figure~\ref{fig:3696area}, we can reach a conclusion that all of the
dOH bonds are located on pore surface accessible from the vacuum--ice interface instead of inside the bulk ASW. On the
surface of pores, a water molecule can form 2, 3, or 4 bonds with
neighbouring water molecules, and they are called 2-, 3-, or 4-coordinated, respectively.
Figure~\ref{fig:wat_doh_area} shows that as long as the ASW is heated to
$\sim$60~K, 2-coordinated water molecules disappear, leaving 3- and 4-
coordinated molecules. Based on the comparison of Figure~\ref{fig:sat_area} with Figure~\ref{fig:wat_doh_area}, we can see that  between 60 K and 140 K, the number of
3-coordinated water molecules is roughly proportional to the total pore surface
area. This indicates that the ratio between 3- and 4-coordinated water molecules
is more or less a constant. As the properties of ASW surface is determined by the relative ratio between 2-, 3-, and 4-coordinated water molecules, this would indicate that other than the differing in the total pore surface area, the
properties of the pore surface remain the same between 60 and 140 K. Previously, \citet{Zubkov2007} performed TPD measurements of N$_2$ adsorbed on both compact and porous ASW, and found that if a correction for surface area is taken into account, the desorption energy distribution on porous ASW is nearly identical to that of compact ASW. Our results based on a different method agrees with their conclusion.
From a laboratory perspective, the independence of surface property on porosity suggests that in some scenarios, one may be
able to carry out experiments on a porous ASW that is annealed to 60 K, and
the result would be the same---aside from a scaling factor--- as that on an ASW that is annealed to 140 K, which has
a compact structure. Since experiments on a porous ASW would in general have
a much higher sensitivity than on the very top surface of a compact ASW, this
conclusion is very useful in laboratory studies of ASW. One example of its
application is presented in \citet{He2018b}, who measured the diffusion of volatile
molecules on the surface of porous ASW. Conclusions in that study would be
applicable to the surface of any ASW (porous or compact) that is annealed to 60
K and above.

The surface of ASW is known to catalyze the formation of molecular species on
dust grains in the ISM. Porous ASW possesses a specific surface area up to
a few hundred m$^2$g$^{-1}$ and therefore may account for most of the catalytic
surface on the dust grains. Even if ASW undergoes thermal processing,
significant residual porosity may be retained \citep{Isokoski2014}. However, it
was unclear how much of the residual porosity can actually contribute to the
catalysis of chemical reactions. The key question here is whether these
remaining pores are closed cavities buried inside the bulk ice, or they are
accessible to volatiles from gas adsorption. Our experimental results in
Figure~\ref{fig:3696area} show that all of the dOH in the ASW can be covered by
CO molecules, which suggests that all of the pore surface area is accessible for
reactive species condensed from the gas phase. The large pore surface actually
contributes to the catalysis of the formation of complex species in the ice.
The fact that all pores are connected all the way to the vacuum--ice interface
suggests the possibility that volatile molecules that are formed on the pore
surface can diffuse and desorb from the ice before the
desorption of water. The desorption of molecules before water desorption has the
potential to explain the observations which found complex organic molecules in
regions with high-extinction \citep{Vasyunin2013,Agundez2015} and regions
outside the water snow line in protoplanetary disks \citep{Oberg2015}.

\section{Conclusions}
\label{sec:conclusions}
In this study we used the infrared absorption spectrum of carbon monoxide as
a tool to measure the pore surface area of amorphous solid water grown by vapor
deposition and annealed at different temperatures. A kinetic Monte Carlo model was used to visualize the porosity and measure the surface area directly. Below are the findings from this study:
\begin{itemize}
  \item Experiental results show that the total pore surface area in 200 ML of ASW at 20 K is equivalent to 46
    ML, and decreases linearly with annealing temperature to $\sim$120 K.
  \item Almost all pores are connected to the vacuum--ice interface and
    accessible for volatiles adsorption.
  \item All dangling OH bonds, as inferred by the 3696 cm$^{-1}$ and 3720
    cm$^{-1}$ features, reside on the surface of pores.
  \item The 3720 cm$^{-1}$ dOH band, which is due to 2-coordinated water
    molecules, disappears when the ASW is heated to 60 K.
  \item The 3696 cm$^{-1}$ dOH band, which is due to 3-coordinated water
    molecules, decreases more or less linearly between $\sim$50
    K and 140 K.
  \item The ratio between 3- and 4-coordinated water molecules on the surface of
    pores remains constant between 60 K and 140 K; this indicates that the
    surface properties, as adsorption of volatiles is concerned, do not change significantly in this temperature range,
    except for the change in the total surface area.
  \item The 2152 cm$^{-1}$ absorption peak observed for CO on ASW is due to the
    interaction of CO with dOH bonds on pore surfaces.
  \item ASW annealed to 60 K or above loses the capability to trap CO molecules
    from the gas phase.
  \item After the first $\sim$10 ML, the fraction of surface molecules to the total number water molecules does not change with thickness.
\end{itemize}

\section{Acknowledgements}
Work at Syracuse University was supported by NSF Astronomy \& Astrophysics
Research Grant number 1615897 to GV. RTG thanks the NASA APRA program for funding through grant number NNX15AG07G.


\begin{thebibliography}{}
\expandafter\ifx\csname natexlab\endcsname\relax\def\natexlab#1{#1}\fi
\providecommand{\url}[1]{\href{#1}{#1}}
\providecommand{\dodoi}[1]{doi:~\href{http://doi.org/#1}{\nolinkurl{#1}}}
\providecommand{\doeprint}[1]{\href{http://ascl.net/#1}{\nolinkurl{http://ascl.net/#1}}}
\providecommand{\doarXiv}[1]{\href{https://arxiv.org/abs/#1}{\nolinkurl{https://arxiv.org/abs/#1}}}

\bibitem[{{Ag{\'u}ndez} {et~al.}(2015){Ag{\'u}ndez}, {Cernicharo}, \&
  {Gu{\'e}lin}}]{Agundez2015}
{Ag{\'u}ndez}, M., {Cernicharo}, J., \& {Gu{\'e}lin}, M. 2015, \aap, 577, L5,
  \dodoi{10.1051/0004-6361/201526317}

\bibitem[{{Alan May} {et~al.}(2013){Alan May}, {Scott Smith}, \&
  {Kay}}]{May2013}
{Alan May}, R., {Scott Smith}, R., \& {Kay}, B.~D. 2013, \jcp, 138, 104502,
  \dodoi{10.1063/1.4793312}

\bibitem[{Bar-nun {et~al.}(1985)Bar-nun, Herman, Laufer, \&
  Rappaport}]{Bar-Nun1985}
Bar-nun, A., Herman, G., Laufer, D., \& Rappaport, M. 1985, Icarus, 63, 317 ,
  \dodoi{https://doi.org/10.1016/0019-1035(85)90048-X}

\bibitem[{{Bieler} {et~al.}(2015){Bieler}, {Altwegg}, {Balsiger}, {Bar-Nun},
  {Berthelier}, {Bochsler}, {Briois}, {Calmonte}, {Combi}, {de Keyser}, {van
  Dishoeck}, {Fiethe}, {Fuselier}, {Gasc}, {Gombosi}, {Hansen}, {H{\"a}ssig},
  {J{\"a}ckel}, {Kopp}, {Korth}, {Le Roy}, {Mall}, {Maggiolo}, {Marty},
  {Mousis}, {Owen}, {R{\`e}me}, {Rubin}, {S{\'e}mon}, {Tzou}, {Waite}, {Walsh},
  \& {Wurz}}]{Bieler2015}
{Bieler}, A., {Altwegg}, K., {Balsiger}, H., {et~al.} 2015, \nat, 526, 678,
  \dodoi{10.1038/nature15707}

\bibitem[{{Bossa} {et~al.}(2012){Bossa}, {Isokoski}, {de Valois}, \&
  {Linnartz}}]{Bossa2012}
{Bossa}, J.~B., {Isokoski}, K., {de Valois}, M.~S., \& {Linnartz}, H. 2012,
  \aap, 545, A82, \dodoi{10.1051/0004-6361/201219340}

\bibitem[{{Bossa} {et~al.}(2014){Bossa}, {Isokoski}, {Paardekooper}, {Bonnin},
  {van der Linden}, {Triemstra}, {Cazaux}, {Tielens}, \&
  {Linnartz}}]{Bossa2014}
{Bossa}, J.~B., {Isokoski}, K., {Paardekooper}, D.~M., {et~al.} 2014, \aap,
  561, A136, \dodoi{10.1051/0004-6361/201322549}

\bibitem[{Brown {et~al.}(1996)Brown, George, Huang, Wong, Rider, Smith, \&
  Kay}]{Brown1996}
Brown, D.~E., George, S.~M., Huang, C., {et~al.} 1996, Journal of Physical
  Chemistry, 100, 4988, \dodoi{10.1021/jp952547j}

\bibitem[{{Bu} {et~al.}(2016){Bu}, {Dukes}, \& {Baragiola}}]{Bu2016}
{Bu}, C., {Dukes}, C.~A., \& {Baragiola}, R.~A. 2016, Applied Physics Letters,
  109, 201902, \dodoi{10.1063/1.4967789}

\bibitem[{{Bu} {et~al.}(2015){Bu}, {Shi}, {Raut}, {Mitchell}, \&
  {Baragiola}}]{Bu2015}
{Bu}, C., {Shi}, J., {Raut}, U., {Mitchell}, E.~H., \& {Baragiola}, R.~A. 2015,
  \jcp, 142, 134702, \dodoi{10.1063/1.4916322}

\bibitem[{{Buch} \& {Devlin}(1991)}]{Buch1991}
{Buch}, V., \& {Devlin}, J.~P. 1991, \jcp, 94, 4091, \dodoi{10.1063/1.460638}

\bibitem[{{Cazaux} {et~al.}(2015){Cazaux}, {Bossa}, {Linnartz}, \&
  {Tielens}}]{Cazaux2015}
{Cazaux}, S., {Bossa}, J.~B., {Linnartz}, H., \& {Tielens}, A.~G.~G.~M. 2015,
  \aap, 573, A16, \dodoi{10.1051/0004-6361/201424466}

\bibitem[{{Clements} {et~al.}(2018){Clements}, {Berk}, {Cooke}, \&
  {Garrod}}]{Clements2018}
{Clements}, A.~R., {Berk}, B., {Cooke}, I.~R., \& {Garrod}, R.~T. 2018,
  Physical Chemistry Chemical Physics, 20, 5553

\bibitem[{{Collings} {et~al.}(2005){Collings}, {Dever}, \&
  {McCoustra}}]{Collings2005}
{Collings}, M.~P., {Dever}, J.~W., \& {McCoustra}, M. R.~S. 2005, Chemical
  Physics Letters, 415, 40, \dodoi{10.1016/j.cplett.2005.08.123}

\bibitem[{{Dartois} {et~al.}(2013){Dartois}, {Ding}, {de Barros}, {Boduch},
  {Brunetto}, {Chabot}, {Domaracka}, {Godard}, {Lv}, {Mej{\'\i}a Guam{\'a}n},
  {Pino}, {Rothard}, {da Silveira}, \& {Thomas}}]{Dartois2013}
{Dartois}, E., {Ding}, J.~J., {de Barros}, A.~L.~F., {et~al.} 2013, \aap, 557,
  A97, \dodoi{10.1051/0004-6361/201321636}

\bibitem[{{Devlin}(1995)}]{Devlin1995}
{Devlin}, J.~P.~and{Buch}, V. 1995, \jcp, 99, 16534

\bibitem[{{Dohn{\'a}lek} {et~al.}(2003){Dohn{\'a}lek}, {Kimmel}, {Ayotte},
  {Smith}, \& {Kay}}]{Dohnalek2003}
{Dohn{\'a}lek}, Z., {Kimmel}, G.~A., {Ayotte}, P., {Smith}, R.~S., \& {Kay},
  B.~D. 2003, \jcp, 118, 364, \dodoi{10.1063/1.1525805}

\bibitem[{{Eldrup} {et~al.}(1985){Eldrup}, {Vehanen}, {Schultz}, \&
  {Lynn}}]{Eldrup1985}
{Eldrup}, M., {Vehanen}, A., {Schultz}, P.~J., \& {Lynn}, K.~G. 1985, Physical
  Review B, 32, 7048, \dodoi{10.1103/PhysRevB.32.7048}

\bibitem[{{Fraser} {et~al.}(2004){Fraser}, {Collings}, {Dever}, \&
  {McCoustra}}]{Fraser2004}
{Fraser}, H.~J., {Collings}, M.~P., {Dever}, J.~W., \& {McCoustra}, M. R.~S.
  2004, \mnras, 353, 59, \dodoi{10.1111/j.1365-2966.2004.08038.x}

\bibitem[{{Garrod}(2013)}]{Garrod2013}
{Garrod}, R.~T. 2013, The Atrophysical Journal, 778, 14

\bibitem[{{Hagen} {et~al.}(1981){Hagen}, {Tielens}, \& {Greenberg}}]{Hagen1981}
{Hagen}, W., {Tielens}, A.~G.~G.~M., \& {Greenberg}, J.~M. 1981, Chemical
  Physics, 56, 367, \dodoi{10.1016/0301-0104(81)80158-9}

\bibitem[{{He} {et~al.}(2018{\natexlab{a}}){He}, {Emtiaz}, {Boogert}, \&
  {Vidali}}]{He2018c}
{He}, J., {Emtiaz}, S., {Boogert}, A., \& {Vidali}, G. 2018{\natexlab{a}},
  \apj, 869, 41, \dodoi{10.3847/1538-4357/aae9dc}

\bibitem[{{He} {et~al.}(2018{\natexlab{b}}){He}, {Emtiaz}, \&
  {Vidali}}]{He2018b}
{He}, J., {Emtiaz}, S., \& {Vidali}, G. 2018{\natexlab{b}}, \apj, 863, 156,
  \dodoi{10.3847/1538-4357/aad227}

\bibitem[{{He} \& {Vidali}(2018)}]{He2018a}
{He}, J., \& {Vidali}, G. 2018, \mnras, 473, 860, \dodoi{10.1093/mnras/stx2412}

\bibitem[{{Herbst} \& {van Dishoeck}(2009)}]{Herbst2009}
{Herbst}, E., \& {van Dishoeck}, E.~F. 2009, Annual Review of Astronomy and
  Astrophysics, 47, 427, \dodoi{10.1146/annurev-astro-082708-101654}

\bibitem[{{Horimoto} {et~al.}(2002){Horimoto}, {Kato}, \&
  {Kawai}}]{Horimoto2002}
{Horimoto}, N., {Kato}, H.~S., \& {Kawai}, M. 2002, \jcp, 116, 4375,
  \dodoi{10.1063/1.1458937}

\bibitem[{{Isokoski} {et~al.}(2014){Isokoski}, {Bossa}, {Triemstra}, \&
  {Linnartz}}]{Isokoski2014}
{Isokoski}, K., {Bossa}, J.~B., {Triemstra}, T., \& {Linnartz}, H. 2014,
  Physical Chemistry Chemical Physics (Incorporating Faraday Transactions), 16,
  3456, \dodoi{10.1039/C3CP54481H}

\bibitem[{{Keane} {et~al.}(2001){Keane}, {Tielens}, {Boogert}, {Schutte}, \&
  {Whittet}}]{Keane2001}
{Keane}, J.~V., {Tielens}, A.~G.~G.~M., {Boogert}, A.~C.~A., {Schutte}, W.~A.,
  \& {Whittet}, D.~C.~B. 2001, \aap, 376, 254,
  \dodoi{10.1051/0004-6361:20010936}

\bibitem[{Kimmel {et~al.}(2001)Kimmel, Dohnálek, Stevenson, Smith, \&
  Kay}]{Kimmel2001b}
Kimmel, G.~A., Dohnálek, Z., Stevenson, K.~P., Smith, R.~S., \& Kay, B.~D.
  2001, The Journal of Chemical Physics, 114, 5295, \dodoi{10.1063/1.1350581}

\bibitem[{{Kimmel} {et~al.}(2001){Kimmel}, {Stevenson}, {Dohn{\'a}lek},
  {Smith}, \& {Kay}}]{Kimmel2001a}
{Kimmel}, G.~A., {Stevenson}, K.~P., {Dohn{\'a}lek}, Z., {Smith}, R.~S., \&
  {Kay}, B.~D. 2001, \jcp, 114, 5284, \dodoi{10.1063/1.1350580}

\bibitem[{{Mitchell} {et~al.}(2017){Mitchell}, {Raut}, {Teolis}, \&
  {Baragiola}}]{Mitchell2017}
{Mitchell}, E.~H., {Raut}, U., {Teolis}, B.~D., \& {Baragiola}, R.~A. 2017,
  \icarus, 285, 291, \dodoi{10.1016/j.icarus.2016.11.004}

\bibitem[{{Mitterdorfer} {et~al.}(2014){Mitterdorfer}, {Bauer}, {Youngs},
  {Bowron}, {Hill}, {Fraser}, {Finney}, \& {Loerting}}]{Mitterdorfer2014}
{Mitterdorfer}, C., {Bauer}, M., {Youngs}, T. G.~A., {et~al.} 2014, Physical
  Chemistry Chemical Physics (Incorporating Faraday Transactions), 16, 16013,
  \dodoi{10.1039/C4CP00593G}

\bibitem[{{{\"O}berg} {et~al.}(2015){{\"O}berg}, {Guzm{\'a}n}, {Furuya}, {Qi},
  {Aikawa}, {Andrews}, {Loomis}, \& {Wilner}}]{Oberg2015}
{{\"O}berg}, K.~I., {Guzm{\'a}n}, V.~V., {Furuya}, K., {et~al.} 2015, \nat,
  520, 198, \dodoi{10.1038/nature14276}

\bibitem[{{Palumbo}(2006)}]{Palumbo2006}
{Palumbo}, M.~E. 2006, \aap, 453, 903, \dodoi{10.1051/0004-6361:20042382}

\bibitem[{{Raut} {et~al.}(2007{\natexlab{a}}){Raut}, {Fam{\'a}}, {Teolis}, \&
  {Baragiola}}]{Raut2007charact}
{Raut}, U., {Fam{\'a}}, M., {Teolis}, B.~D., \& {Baragiola}, R.~A.
  2007{\natexlab{a}}, \jcp, 127, 204713, \dodoi{10.1063/1.2796166}

\bibitem[{{Raut} {et~al.}(2007{\natexlab{b}}){Raut}, {Teolis}, {Loeffler},
  {Vidal}, {Fam{\'a}}, \& {Baragiola}}]{Raut2007compaction}
{Raut}, U., {Teolis}, B.~D., {Loeffler}, M.~J., {et~al.} 2007{\natexlab{b}},
  \jcp, 126, 244511, \dodoi{10.1063/1.2746858}

\bibitem[{{Schmitt} {et~al.}(1989){Schmitt}, {Greenberg}, \&
  {Grim}}]{Schmitt1989}
{Schmitt}, B., {Greenberg}, J.~M., \& {Grim}, R.~J.~A. 1989, \apj, 340, L33,
  \dodoi{10.1086/185432}

\bibitem[{Smith {et~al.}(1997)Smith, Huang, Wong, \& Kay}]{Smith1997}
Smith, R.~S., Huang, C., Wong, E. K.~L., \& Kay, B.~D. 1997, Phys. Rev. Lett.,
  79, 909, \dodoi{10.1103/PhysRevLett.79.909}

\bibitem[{Smith {et~al.}(2009)Smith, Zubkov, Dohnálek, \& Kay}]{Smith2009}
Smith, R.~S., Zubkov, T., Dohnálek, Z., \& Kay, B.~D. 2009,
  \dodoi{10.1021/jp804902p}

\bibitem[{{Stevenson} {et~al.}(1999){Stevenson}, {Kimmel}, {Dohnalek}, {Smith},
  \& {Kay}}]{Stevenson1999}
{Stevenson}, K.~P., {Kimmel}, G.~A., {Dohnalek}, Z., {Smith}, R.~S., \& {Kay},
  B.~D. 1999, Science, 283, 1505, \dodoi{10.1126/science.283.5407.1505}

\bibitem[{{Vastel} {et~al.}(2014){Vastel}, {Ceccarelli}, {Lefloch}, \&
  {Bachiller}}]{Vastel2014}
{Vastel}, C., {Ceccarelli}, C., {Lefloch}, B., \& {Bachiller}, R. 2014, \apj,
  795, L2, \dodoi{10.1088/2041-8205/795/1/L2}

\bibitem[{{Vasyunin} \& {Herbst}(2013)}]{Vasyunin2013}
{Vasyunin}, A.~I., \& {Herbst}, E. 2013, \apj, 769, 34,
  \dodoi{10.1088/0004-637X/769/1/34}

\bibitem[{{Zheng} {et~al.}(2007){Zheng}, {Jewitt}, \& {Kaiser}}]{Zheng2007}
{Zheng}, W., {Jewitt}, D., \& {Kaiser}, R.~I. 2007, Chemical Physics Letters,
  435, 289, \dodoi{10.1016/j.cplett.2007.01.013}

\bibitem[{Zubkov {et~al.}(2007)Zubkov, Smith, Engstrom, \& Kay}]{Zubkov2007}
Zubkov, T., Smith, R.~S., Engstrom, T.~R., \& Kay, B.~D. 2007, J. Chem. Phys.,
  127, 184708, \dodoi{10.1063/1.2790433}

\end{thebibliography}
\end{document}